\documentclass[english,twocolumn,superscriptaddress,showpacs,floats]{revtex4}
\usepackage{epsfig,epsf,amsmath,amsthm,amsfonts,
            graphics,graphicx,amssymb,babel,subfigure}

\usepackage{hyperref}

\newcommand{\sings}{{\it single-sided}}
\newcommand{\inc}{{\it inclusive}}

\newcommand{\ffake}{f_{\xi<5\times 10^{-6}}}

\begin{document}
\preprint{CERN-PH-EP-2011-047}
%
% ======> Title of the paper goes here <====================
%
\title{Measurement of the Inelastic Proton-Proton Cross-Section at $\sqrt{s}=7$~
  TeV with the ATLAS Detector}

%
% ==============> Text of the abstract goes here <=====================
% 
\begin{abstract}
 A first measurement of the inelastic cross-section is presented for
proton-proton collisions at a center of mass energy $\sqrt{s}=7$~TeV
using the ATLAS detector at the Large Hadron Collider. In a dataset corresponding to an integrated luminosity of 20~$\mu$b$^{-1}$, events are selected by requiring hits on scintillation counters mounted in the forward region of the detector.  An inelastic cross-section of $60.3 \pm 2.1$~mb is measured for $\xi > 5\times 10^{-6}$, where $\xi=M_X^2/s$ is calculated from the
invariant mass, $M_X$, of hadrons selected using the largest rapidity
gap in the event. For diffractive events this corresponds to requiring
at least one of the dissociation masses to be larger than 15.7~GeV.

\end{abstract}
\maketitle

Since the earliest days of particle physics, measurements of the total $pp$ and $p\bar{p}$ cross-sections and their theoretical understanding have been topics of much interest~\cite{pdg}. The cross-sections can not yet be calculated by quantum chromodynamics (QCD), and many approaches have been used to describe the existing measurements. General arguments based on unitarity, analyticity, and factorisation imply a bound (the Froissart bound~\cite{froissart,martin1}) on the high-energy behaviour of total hadronic cross-sections. This bound is independent of the details of the strong interaction dynamics and states that the total cross-section can not rise faster than $\ln^2(s)$, where $\sqrt{s}$ is the centre-of-mass energy. Recently it has been extended to the inelastic cross-section~\cite{martin2}. Existing experimental data~\cite{pdg} show a rise in the hadronic cross-sections with $s$, but it is unclear whether the asymptotic behaviour has already been reached. With the data presented in this letter we shed further light on the high energy behaviour of the inelastic cross-section.

The most common models that describe the data up to $\sqrt{s} = 1.8$~TeV predict a rise of the cross-section with a simple power law ($s^{\alpha(0)-1}$ where $\alpha(0)$ denotes the Pomeron-trajectory intercept)~\cite{regge,dl1,dl2} or as a logarithm~\cite{eikonqcd,blockCahn,block,cudell}. Others employ QCD for aspects of the calculation~\cite{kmr2011,glm,godbole}. However, while the phenomenological description of the existing data is largely adequate, there are significant uncertainties on the extrapolation to higher energies, partly due to a long-standing $2.7\sigma$ discrepancy between the two highest energy collider measurements of the total $p\bar{p}$ cross-section by CDF~\cite{cdf} and by E811~\cite{e811}. 

In this letter a measurement of the inelastic $pp$ cross-section is presented using data taken by the ATLAS experiment~\cite{atlas} at the Large Hadron Collider (LHC)~\cite{lhc} at $\sqrt{s}=7$~TeV. The data considered were collected during a single 8-hour fill beginning March 31st, 2010, corresponding to an integrated luminosity of $20.3 \pm 0.7$~$\mu$b$^{-1}$ and a peak instantaneous luminosity of $1.2 \times 10^{27}$~cm$^{-2}$s$^{-1}$. The mean number of interactions per crossing in this fill is approximately 0.01.  The analysis uses highly efficient scintillation counters to detect inelastic collisions. They are insensitive to diffractive dissociation processes in which the dissociation systems have small invariant masses, $M_X$.  Their acceptance corresponds approximately to $\xi = M_X^2 / s > 5 \times 10^{-6}$, equivalent to $M_X>15.7$~GeV for $\sqrt{s}=7$~TeV.  
The cross-section measurement presented here is restricted to this kinematic range. However, in order to compare the data with previous measurements, an extrapolation of the cross-section is performed to the full $\xi$ range, $\xi>m_p^2/s$ where $m_p$ is the proton mass. 

The ATLAS detector is described in detail elsewhere~\cite{atlas}. The beam-line is surrounded by a tracking detector that uses silicon pixel, silicon strip, and straw tube technologies and is embedded in a 2~T magnetic field. The tracking system covers the pseudorapidity~\cite{coord} range $|\eta|<2.5$. 
It is surrounded by electromagnetic and hadronic calorimeters covering $|\eta|< 3.2$, which are complemented by a forward hadronic calorimeter covering $3.1<|\eta|<4.9$.  Minimum Bias Trigger Scintillator (MBTS) detectors, the primary detectors used in this measurement, are mounted in front of the endcap calorimeters on both sides of the interaction point at $z=\pm 3.56$~m and cover the range $2.09<|\eta|<3.84$. Each side consists of 16 independent counters divided into two rings; the inner 8 counters cover the rapidity range $2.83<|\eta|<3.84$ and the outer 8 counters cover the range $2.09<|\eta|<2.83$. Each individual counter spans 45$^\circ$ of the azimuthal angle ($\phi$), and 31 out of 32 counters were operational. 
The luminosity is measured using a Cherenkov light detector, LUCID, which is located at $z = \pm17$~m. The luminosity calibration has been determined during dedicated van der Meer beam scans  
to a precision of 3.4\%~\cite{vdm,lumipaper}. 

Monte Carlo (MC) simulations are used to determine the acceptance of the event selection and to assess systematic uncertainties. 
The detector response to the generated events is simulated using the ATLAS simulation~\cite{atlassim} based on {\sc Geant4}~\cite{g4}, and both the simulated and data events are reconstructed and analysed with the same software. The {\sc Pythia6}~\cite{py6}, {\sc Pythia8}~\cite{py8} and {\sc Phojet}~\cite{phojet} generators are used to predict properties of inelastic collisions. These generators distinguish between different processes that contribute to inelastic $pp$ interactions: {\it single dissociative} (SD) processes, $pp \rightarrow pX$, in which one proton dissociates; {\it double dissociative} (DD) processes, $pp \rightarrow XY$, in which both protons dissociate with no net color flow between the systems $X$ and $Y$; and {\it non-diffractive} (ND) processes in which color flow is present between the two initial-state protons. The
model by Schuler and Sj\"ostrand~\cite{sjoestrand94}, used by  {\sc Pythia6} and {\sc Pythia8}, predicts cross-sections of 48.5~mb, 13.7~mb and 9.3~mb for the ND, SD and DD processes, respectively. While the cross-sections used by {\sc Pythia6} and {\sc Pythia8} are identical, they differ in the modelling of the hadronic final state. {\sc Phojet} predicts the corresponding cross-sections as 61.6~mb (ND), 10.7~mb (SD) and 3.9~mb (DD).  Due to differences in implementation of the interface between large $\xi$ diffractive (SD and DD) processes and ND processes in {\sc Pythia} and {\sc Phojet} the fractional contribution of these processes is a model dependent quantity.  {\sc Phojet} also includes a 1.1~mb contribution from {\it central diffraction} (CD), $pp \rightarrow ppX$, a process not implemented in {\sc Pythia}, wherein neither proton dissociates but the Pomeron-trajectory exchange results in energy loss for the protons and the production of a central system of particles.  
The MC generators define the inelastic cross-section as the sum of these contributions, and thus Schuler and Sj\"ostrand ({\sc Phojet}) predicts an inelastic cross-section of 71.5~mb (77.3~mb). Other recent predictions for this cross-section at~$\sqrt{s} = 7$~TeV are 69~mb~\cite{block}, 65-67~mb~\cite{kmr2011}, 68~mb~\cite{glm} and 60-75~mb~\cite{godbole}. 

The variable $\xi$ is defined at the particle level by dividing the final state particles into two systems, X and Y. The mean $\eta$ of the two particles separated by the largest pseudorapidity gap in the event is used to assign all particles with greater pseudorapidity to one system and all particles with smaller pseudorapidity to the other~\cite{refmx}.  The mass, $M_{X,Y}$, of each system is calculated and the higher mass system is defined as $X$ while the lower mass system is defined as $Y$. The variable $\xi$ is then given by $\xi=M_X^2/s$ and it is bounded by the elastic limit of $\xi > m_p^2/s$. 
Due to the limited MBTS detector acceptance in $\eta$, this measurement is restricted to the range $\xi > 5 \times 10^{-6}$; there is no restriction on $M_Y$.  Several models are used for the dependence of the diffractive cross-sections on $\xi$.  The Schuler and Sj\"ostrand model has a relatively flat dependence on $\xi$, while the {\sc Phojet} model predicts a slight decrease with decreasing $\xi$. {\sc Pythia8} has several additional predictions for the $\xi$-dependence of the diffractive cross-sections which are considered. Bruni and Ingelman~\cite{bri} predict a flat $\xi$-dependence while Donnachie and Landshoff (DL)~\cite{dlmodel}, Berger {\it et al.}~\cite{berger}, and Streng~\cite{streng} predict 
\begin{eqnarray*}
\frac{{\rm d}\sigma_{SD}}{{\rm d}\xi} & \propto& \frac{1}{\xi^{1+\epsilon}}\left(1+\xi\right)
\end{eqnarray*}
where $\epsilon=\alpha(0)-1$.
Values of $\epsilon$ between $0.06$ and $0.10$, and of $\alpha'$ between $0.10$ and $0.40$~GeV$^{-2}$ are considered for the DL model.  $\alpha'$ is the slope of the Pomeron trajectory which is assumed to be linear such that $\alpha(t) = \alpha(0) + \alpha^\prime t$.
The DL model with $\epsilon=0.085$ and $\alpha' = 0.25$~GeV$^{-2}$ with {\sc Pythia8} fragmentation is the default model in this analysis and the other models are used to assess uncertainties in the modelling of diffractive events.

Experimentally the cross-section is calculated using 
\begin{eqnarray*}
\sigma_{inel}(\xi>5 \times 10^{-6}) & = & \frac{(N-N_{BG})}{\epsilon_{trig} \times\int L{\rm d}t}  \times \frac{1-\ffake}{\epsilon_{sel}}
\end{eqnarray*}
where $N$ is the number of selected events, $N_{BG}$ is the number of background events, $\ffake$ is the fraction of events that pass the event selection but have $\xi<5 \times 10^{-6}$, $\int L{\rm d}t$ is the integrated luminosity, and $\epsilon_{trig}$ and $\epsilon_{sel}$ are the trigger and offline event selection efficiencies in the selected $\xi$-range. Note that in this analysis $\xi$ is defined only at the particle level; the cut value at $5\times10^{-6}$ was chosen such that the efficiency of the MBTS requirement is greater than 50\% for any $\xi$-values greater than $5\times10^{-6}$.  In this measurement $N_{BG}$ and $\epsilon_{trig}$ are determined directly from the data. The MBTS individual counter efficiencies in the MC simulation are tuned to match the observed efficiencies in data.  Then $\epsilon_{sel}$ and $\ffake$ are taken from the tuned MC simulation. In order to reduce the uncertainties in the factors taken from MC simulation, the relative diffractive dissociation cross-section, $f_{D} = \frac{\sigma_{SD}+\sigma_{DD}+\sigma_{CD}}{\sigma_{inel}}$ for each generator is constrained. Each of these steps is described in detail below.

The MBTS functions as a trigger by determining the number of scintillation counters with a signal passing a leading-edge discriminator; in this analysis at least one trigger signal must be present. In the offline reconstruction, the MBTS signals are fit to obtain the total charge and timing of the signal.  The offline event selection requires at least two counters with a charge larger than $0.15$~pC.  This threshold is set to be well above the noise level, which is well described by a Gaussian centred at zero of width 0.02~pC.  This {\it inclusive} sample contains 1,220,743 data events. In order to constrain the diffractive components a subset of events is selected, the {\it single-sided} sample, which contains events that have at least two hits on one side of the MBTS detector and no hits on the opposing side (in $z$).  In the data 122,490 single-sided events are observed.

Backgrounds arise from beam-related interactions, such as collisions of the beam with gas particles in the beam-pipe or with material upstream from the detector, and slowly-decaying, collision-induced radiation termed ``afterglow''~\cite{lumipaper}.  Additionally, instrumental noise and cosmic rays provide backgrounds which were studied and found to be negligible for this analysis. 
The beam-related backgrounds are determined using the number of selected events collected in this fill with the non-colliding bunches, i.e. when only one proton bunch was passing through ATLAS~\cite{minbiaspaper}, normalised by the ratio of the number of protons in the colliding to the non-colliding bunches. The single-sided selection contains $422\pm 28$ background events and the inclusive sample contains $N_{BG}=1,574 \pm 54$ background events, corresponding to 0.3\% and 0.1\% of the total samples, respectively. 
In addition, there is an in-time afterglow component due to the scattering of secondary low-energy particles produced in the same collision event which can give additional hits, causing low-activity events to migrate into the selected event sample. This contribution is evaluated to be at most 0.4\% for  the inclusive, and 3.6\%  for the single-sided, samples by examining the asymmetry of the absolute timing measurement of the MBTS counters. We conservatively assume a 100\% uncertainty on both background sources which covers any residual impact of the afterglow on the background subtraction, any uncertainty in the beam current measurements and the uncertainty due to in-time afterglow. The resulting overall uncertainty on the number of background events $N_{BG}$ is given by the quadratic sum of the two components and is 0.4\%.

The trigger efficiency of the MBTS detector with respect to the offline requirement, $\epsilon_{trig}$, is measured to be $99.98^{+0.02}_{-0.12}\%$ (statistical errors) using events triggered randomly on colliding beams.  The systematic uncertainty on $\epsilon_{trig}$ is determined using a second, independent trigger as reference.  The difference between the two efficiency determinations leads to a 0.1\% uncertainty on the cross-section measurement.  

The data and MC simulation agreement in the MBTS counter response is checked using other detector subsystems with overlapping $\eta$ ranges:
charged particles reconstructed by the tracking detector ($2.09<|\eta|<2.5$), and calorimeter showers in the inner wheel of the 
electromagnetic calorimeter ($2.5<|\eta|<3.2$) and in the forward calorimeter ($3.1<|\eta|<3.84$).  The efficiency with respect to a track (calorimeter energy deposit) to have a signal above the 0.15~pC threshold in the outer (inner) counters is on average 98.5\% (97.5\%) for the data and a constant 99.4\% (98.7\%) in the MC simulation. The individual counter efficiencies deviate by up to 2.0\% (2.5\%) from the average in the data. The MC simulation is corrected to match the data efficiency and the maximum variations in the counter responses are considered as a systematic uncertainty. This results in a 0.1\% uncertainty on the cross-section measurement. 

The offline selection efficiency, $\epsilon_{sel}$, depends on the amount of material traversed by particles before hitting the MBTS detector.  The rate of photons (primarily from $\pi^0$ decays) converting to electrons which are subsequently detected by the MBTS increases with additional material, resulting in an increase of $\epsilon_{sel}$.  
%The dominant effect arises from photons (primarily from $\pi^0$ decays) converting to electrons which are subsequently detected by the MBTS. Thus the efficiency for an event to pass the selection increases with increasing material. 
Second order effects arise from charged particles scattering out of the MBTS acceptance region (decreasing $\epsilon_{sel}$), or charged particles scattering into the acceptance region (increasing $\ffake$). %which has been modeled accurately using engineering drawings. 
Within the tracking volume ($|\eta|<2.5$) the material distribution has been studied using conversion electrons and $K^0_s \to \pi^+\pi^-$ decays, and is known to better than  $\pm 5$\% in the central region of the detector and to $\pm 30$\% for $2.2<|\eta|<2.5$~\cite{minbiaspaper}. In the region $|\eta|>2.5$ the material is dominated by the cooling and electrical services to the silicon pixel detector, and an uncertainty of $\pm 40$\% is assumed. This is validated {\it in-situ} using the fraction of events where we observe significant energy in the forward calorimeters but no signal (above noise) in the MBTS detector. The resulting systematic uncertainty on the cross-section is $0.2\%$.

Misalignments of the MBTS detector with respect to the nominal centre of the detector could change the event selection efficiency for a particular value of $\xi$.  Misalignments of up to 10~mm were considered and found to have a negligible impact. A misalignment of 10~mm is conservative compared to the survey precision and any known misalignments within the ATLAS experiment~\cite{atlas}.

The fractional contribution of diffractive events, $f_D$, is constrained by the ratio of single-sided to inclusive events, $R_{ss}$. The MC generators predict that less than 1\% of the ND process pass the single-sided event selection, while $27-41\%$ of the SD and DD processes pass the single-sided selection. For all models the inclusive sample is dominated by ND events, therefore the ratio of single-sided to inclusive events is sensitive to the relative fraction of diffractive events.

The measured $R_{ss}$ in the data is
$R_{ss}= [10.02 \pm 0.03 {\rm (stat.)}~^{+0.1}_{-0.4} {\rm (syst.)}] \%$, where the systematic error includes the uncertainties on the backgrounds, the MBTS response and the material.

\begin{figure}[htb]
\begin{center}
\includegraphics[width=0.5\textwidth]{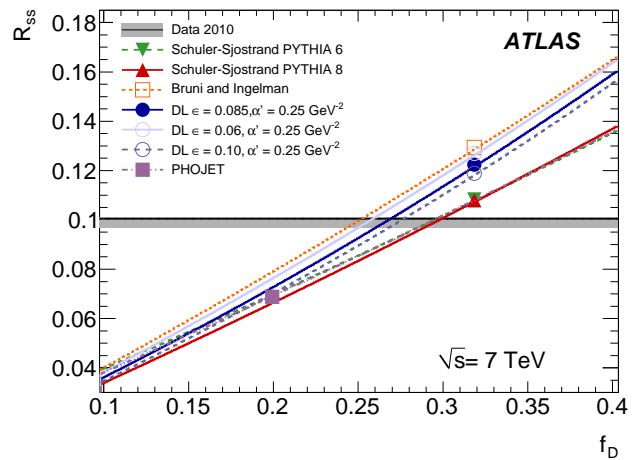}
\caption{The ratio of the single-sided to inclusive event sample $R_{ss}$ as a function of the fractional contribution of diffractive events to the inelastic cross-section $f_D$. The data value for $R_{ss}$ is shown as the horizontal line with its systematic uncertainties (grey band). Also shown are predictions of several models as a function of an assumed value of $f_D$. The default $f_D$ value (32.2\% for all models but {\sc Phojet} which is 20.2\%) is indicated by the markers. 
}
\label{fig:rss}
\end{center}
\end{figure}

Fig.~\ref{fig:rss} compares the observed value of $R_{ss}$ to the predictions of several models as a function of $f_D$. 
The intersection of the $R_{ss}$ value measured in data with the prediction is used as the central value of $f_D$ for each model. The systematic uncertainty on $f_D$ is determined by the maximum and minimum values consistent with the $1\sigma$ uncertainty on the data when varying the double- to single-dissociation event ratio between 0 and 1. The resulting value using the default DL model 
is $f_D=26.9^{+2.5}_{-1.0} \%$.

\begin{figure}[htb]
\begin{center}
\subfigure[]{\includegraphics[width=0.5\textwidth]{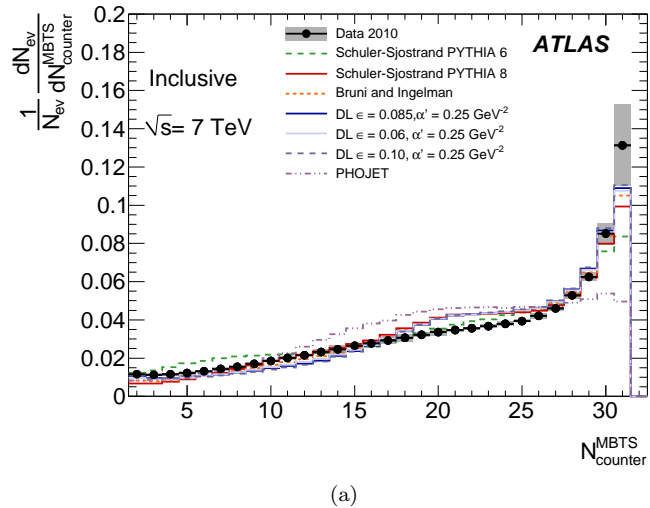}\label{mbtshitA}}
\subfigure[]{\includegraphics[width=0.5\textwidth]{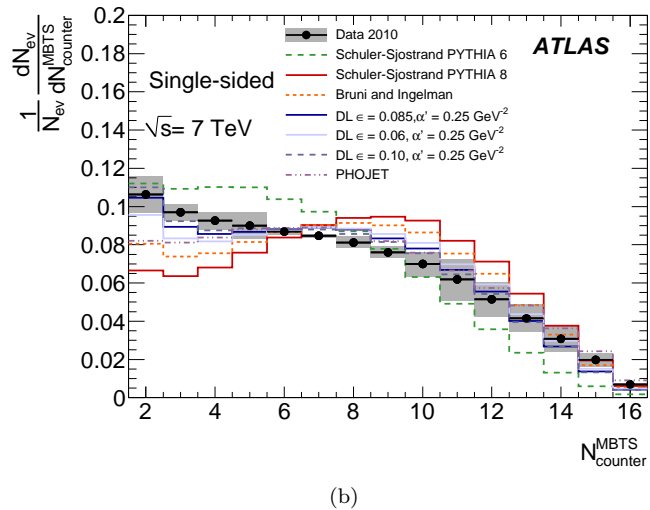}\label{mbtshitB}}
\caption{The MBTS multiplicity distribution in the data (filled circles) compared with MC expectations for the \inc~\subref{mbtshitA} and \sings~\subref{mbtshitB} samples for several MC models (histograms) using the fitted $f_D$ values. The band around the data indicates the systematic uncertainty due to the MBTS detector response and the amount of material in front of the MBTS detector.}
\label{fig:mbtshit}
\end{center}
\end{figure}

The acceptance calculation relies on the MC generators to provide an adequate description of the particle multiplicity in the acceptance region. The validity of the MC description is assessed by examining the hit multiplicity in the MBTS detector in the inclusive and single-sided event samples as shown in Fig.~\ref{fig:mbtshit}. While none of the generators gives a perfect description, the data lie between the models at low multiplicity which is most important for the measurement.  The default DL model describes the single-sided sample well, giving confidence in the diffractive modelling.  We use the difference in the MC correction factor determined with {\sc Pythia8} and {\sc Pythia6} as the uncertainty due to the fragmentation model, leading to a 0.4\% uncertainty on the cross-section.  The maximum difference between the default DL model and all other models is taken as the uncertainty due to the underlying $\xi$ distribution.  Variations of $\alpha'$ have a negligible effect on the acceptance. Among all the models considered, the {\sc Phojet} model gives the largest difference in the correction factor, leading to a 0.4\% uncertainty on the cross-section. 

The final result for the measured inelastic cross-section is calculated using the default DL model of $\epsilon =0.085$ and $\alpha'=0.25$, 
which yields $f_D=26.9\%$, $\epsilon_{sel}=98.77\%$, and $\ffake=0.96\%$. Together with $\epsilon_{trig}=99.98\%$, $N=1,220,743$, $N_{BG}=1,574$ and $\int L{\rm d}t=20.25$~$\mu$b$^{-1}$ this results in 
$\sigma_{inel}(\xi > 5 \times 10^{-6}) = 60.3 \pm 0.05 {\rm (stat)} \pm 0.5 {\rm (syst)} \pm 2.1 {\rm (lumi)~mb}.$
The systematic uncertainty includes all contributions discussed above and listed in Table~\ref{tab:sys}; the dominant uncertainty arises from the luminosity calibration and is quoted separately.
\begin{table}
\begin{center}
\begin{tabular}{|l| c |}
\hline
Source &Uncertainty (\%)\\ \hline
Trigger Efficiency & 0.1  \\
MBTS Response & 0.1 \\
Beam Background & 0.4 \\ 
$f_{D}$ & 0.3 \\
MC Multiplicity &0.4 \\
$\xi$-Distribution & 0.4\\ 
Material & 0.2 \\ 
Luminosity & 3.4  \\ \hline
Total & 3.5 \\
\hline
\end{tabular}
\caption{\label{tab:sys} Sources of systematic uncertainty and their effect on the cross-section measurement.}
\end{center}
\end{table}

\begin{figure}[htbp]
\begin{center}
\includegraphics[width=0.5\textwidth]{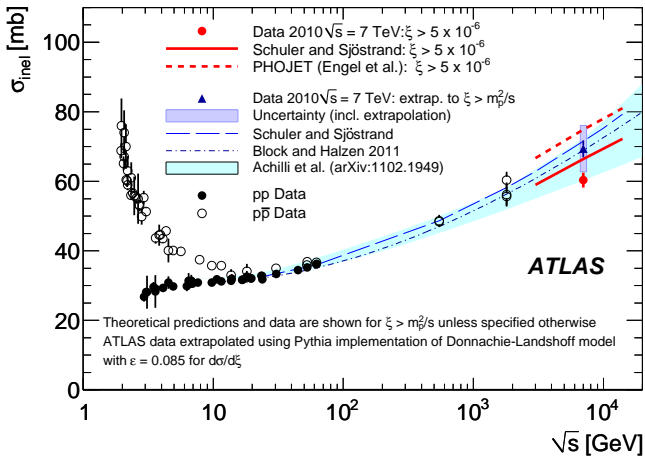}
\caption{The inelastic cross-section versus $\sqrt{s}$. The ATLAS measurement for $\xi > 5 \times 10^{-6}$ is shown as the red filled circle and compared with the predictions of Schuler and Sj\"ostrand and {\sc Phojet} for the same phase space.  Data (filled circles for $pp$ data and unfilled circles for $p\bar{p}$ data) from several experiments are compared with the predictions of the $pp$ inelastic cross-section from Schuler and Sj\"ostrand~\cite{sjoestrand94} (as used by {\sc Pythia}), by Block and Halzen~\cite{block} and by Achilli {\it et al.}~\cite{godbole}. An extrapolation from the measured range of $\xi>5 \times 10^{-6}$ to the full inelastic cross-section using the acceptance of $87\pm 10\%$ is also shown (blue filled triangle).  The experimental uncertainty is indicated by the error bar while the total (including the extrapolation uncertainty) is represented by the blue shaded area.} %The value is artificially shifted to a slightly larger $\sqrt{s}$ value for display purposes.}
\label{fig:totxsec}
\end{center}
\end{figure}

The measurement is compared to the predictions in Figure~\ref{fig:totxsec} and Table~\ref{tab:results}. The predictions by the Schuler-Sj\"ostrand model~(66.4~mb) and the {\sc Phojet} model~(74.2~mb) are both higher than the data. The prediction of 51.8-56.2~mb by Ryskin~{\it et al.}~\cite{kmr2011} is slightly lower than the data.

In order to compare with previous measurements and analytic models, the fractional contribution to the inelastic cross-section of events passing the $\xi>5 \times 10^{-6}$ cut is determined from the models and used to extrapolate the measurement to the full inelastic cross-section.  This fraction is 87.3\% for the default model of DL with $\epsilon = 0.085$ and $\alpha'=0.25$.  The other models considered give fractions ranging from 96\%({\sc Phojet}) to 86\%(DL with $\epsilon = 0.10$). Recent calculations also yield values between 79\% and 84\%~\cite{kmr2011}.  Thus 87\% is taken as the default value for this fraction and an uncertainty of 10\% is taken due to the extrapolation uncertainty on the $\xi$-dependence. The resulting inelastic cross-section value is $\sigma_{inel} = 69.4 \pm 2.4 {\rm (exp.)} \pm 6.9 {\rm (extr.)~mb}$ where exp includes the statistical and experimental systematic errors, including the luminosity uncertainty.

This result is shown in Fig.~3 compared to several theoretical predictions and a variety of data at lower $\sqrt{s}$. 
The measurement within the kinematic range $\xi>5 \times 10^{-6}$ is significantly lower than the predictions of Schuler and Sj\"ostrand and  {\sc Phojet}. The extrapolated value agrees within the large extrapolation uncertainty with the predictions from {\sc Pythia}, which uses a power law dependence on $\sqrt{s}$. It also agrees with Block and Halzen~\cite{block} (which has a logarithmic $\sqrt{s}$ dependence), and with other recent theoretical predictions that vary between 60 and 72~mb~\cite{kmr2011,glm,godbole}. It should be stressed that this extrapolation relies on the prediction of the $\xi$-dependence of the cross-section.

The measurement and a variety of theoretical predictions are also summarised in Table~\ref{tab:results}. 
\begin{table}
\begin{center}
\begin{tabular}{|l| c |}
\hline
\multicolumn{2}{|c|}{$\sigma(\xi>5 \times 10^{-6}$) [mb]} \\ \hline
ATLAS Data 2010 &  $60.33 \pm 2.10 {\rm (exp.)}$\\
Schuler and Sj\"ostrand & $66.4$ \\
{\sc Phojet} & $74.2$ \\
Ryskin {\it et al.} & $51.8-56.2$\\
\hline\hline
\multicolumn{2}{|c|}{$\sigma(\xi>m_p^2/s)$ [mb]} \\ \hline
ATLAS Data 2010& $69.4 \pm 2.4 {\rm (exp.)} \pm 6.9 {\rm (extr.)}$\\
Schuler and Sj\"ostrand & $71.5$ \\
{\sc Phojet} & $77.3$ \\
Block and Halzen & $69$ \\
Ryskin {\it et al.} & $65.2-67.1$\\
Gotsman {\it et al.} & $68$ \\
Achilli {\it et al.} & $60-75$ \\
\hline
\end{tabular}
\caption{\label{tab:results} Measurement and theoretical predictions of the inelastic cross-section for the restricted kinematic range, $\xi>5\times 10^{-6}$, and for the full kinematic range, $\xi>m_p^2/s$. The experimental uncertainty (exp.) includes the statistical, systematic and luminosity uncertainties. The extrapolation uncertainty (extr.) only applies to the full kinematic range and is listed separately.}
\end{center}
\end{table}

In conclusion, a first measurement of the inelastic cross-section has been presented for $pp$ collisions at $\sqrt{s}=7$~TeV with a precision of 3.5\%.  The measurement is limited to the kinematic range corresponding to the detector acceptance: $\xi > 5 \times 10^{-6}$.  Phenomenological predictions for both a power law dependence and a logarithmic rise of the cross-section with energy
are consistent with the measurement.

\section*{Acknowledgements}

We wish to thank CERN for the efficient commissioning and operation of the LHC during this initial high-energy data-taking period as well as the support staff from our institutions without whom ATLAS could not be operated efficiently.

We acknowledge the support of ANPCyT, Argentina; YerPhI, Armenia; ARC, Australia; BMWF, Austria; ANAS, Azerbaijan; SSTC, Belarus; CNPq and FAPESP, Brazil; NSERC, NRC and CFI, Canada; CERN; CONICYT, Chile; CAS, MOST and NSFC, China; COLCIENCIAS, Colombia; MSMT CR, MPO CR and VSC CR, Czech Republic; DNRF, DNSRC and Lundbeck Foundation, Denmark; ARTEMIS, European Union; IN2P3-CNRS, CEA-DSM/IRFU, France; GNAS, Georgia; BMBF, DFG, HGF, MPG and AvH Foundation, Germany; GSRT, Greece; ISF, MINERVA, GIF, DIP and Benoziyo Center, Israel; INFN, Italy; MEXT and JSPS, Japan; CNRST, Morocco; FOM and NWO, Netherlands; RCN, Norway;  MNiSW, Poland; GRICES and FCT, Portugal;  MERYS (MECTS), Romania;  MES of Russia and ROSATOM, Russian Federation; JINR; MSTD, Serbia; MSSR, Slovakia; ARRS and MVZT, Slovenia; DST/NRF, South Africa; MICINN, Spain; SRC and Wallenberg Foundation, Sweden; SER,  SNSF and Cantons of Bern and Geneva, Switzerland;  NSC, Taiwan; TAEK, Turkey; STFC, the Royal Society and Leverhulme Trust, United Kingdom; DOE and NSF, United States of America.  

The crucial computing support from all WLCG partners is acknowledged gratefully, in particular from CERN and the ATLAS Tier-1 facilities at TRIUMF (Canada), NDGF (Denmark, Norway, Sweden), CC-IN2P3 (France), KIT/GridKA (Germany), INFN-CNAF (Italy), NL-T1 (Netherlands), PIC (Spain), ASGC (Taiwan), RAL (UK) and BNL (USA) and in the Tier-2 facilities worldwide.


\begin{thebibliography}{99}
\bibitem{coord} ATLAS uses a right-handed coordinate system with its origin at the nominal interaction point (IP) in the centre of the detector and the $z$-axis coinciding with the axis of the beam pipe. The $x$-axis points from the IP to the centre of the LHC ring, and the $y$ axis points upward. Cylindrical coordinates $(r,\phi)$ are used in the transverse plane, $\phi$ being the azimuthal angle around the beam pipe. The pseudorapidity is defined in terms of the polar angle $\theta$ as $\eta=-\ln\tan(\theta/2)$.
\end{thebibliography}
\end{document}